# Effect of meditation on scaling behavior and complexity of human heart rate variability

A. Sarkar and P. Barat

Variable Energy Cyclotron Centre

1/AF Bidhan Nagar, Kolkata 700064, India

## Abstract

The heart beat data recorded from samples before and during meditation are analyzed using two different scaling analysis methods. These analyses revealed that mediation severely affects the long range correlation of heart beat of a normal heart. Moreover, it is found that meditation induces periodic behavior in the heart beat. The complexity of the heart rate variability is quantified using multiscale entropy analysis and recurrence analysis. The complexity of the heart beat during mediation is found to be more.

## Introduction

Physiological signals are endpoint manifestation of cellular events that depict how physiological systems vary over time. Extraordinary complexity of the physiological signals and the recognition that physiologic time series contain hidden information has stimulated growing interest in applying concepts and techniques of statistical and nonlinear physics to a wide range of biomedical problem. The human heart beat is one of the important examples of complex physiologic fluctuations. It is considered to be controlled by the autonomic nervous system, which also controls many other vital functions within the body. This system is divided into two parts, the sympathetic and the parasympathetic. The sympathetic nervous system helps to mobilize the body for action. When a person is under stress, it produces 'fight-or-flight' response. The parasympathetic nervous system creates what is usually called the 'rest and digest' response. The responses of the parasympathetic system oppose those of the sympathetic system. The nonlinear interaction between these two branches of the autonomic nervous system and the competition between their stimulation effects are postulated mechanisms for erratic behavior of heart beat. Thus, heart beat is considered as a suitable marker for the estimation of the autonomic nervous system function.

Meditation is considered to be one of the complementary and alternative medicines to cardiovascular disease. People practice meditation for a number of health related purposes. However, it is not fully known what changes occur in body during meditation. While scientists are studying whether meditation may afford meaningful health benefits, they are also looking at how it may do so. One way some types of meditation might work is by reducing the activity of the sympathetic nervous system and increasing the activity of the parasympathetic nervous system.

Peng *et. al.* [1] first made a quantitative study on the effect of meditation on human heart beat. They collected and analyzed continuous heart rate time series from two groups of healthy young and adults before and during two well known form of meditation. Their intention was to determine (i) whether there are any distinctive heart rate dynamics during the meditation practices and (ii) whether such meditative states induce a quiescent (less variable) or active (more variable) pattern of autonomic response. They applied spectral analysis and a technique based on Hilbert transform to quantify the heart rate dynamics during meditation. They reported extremely prominent heart rate oscillations associated with slow breathing during meditation. The amplitude of oscillations during meditation is significantly greater than in the pre-meditation control state.

Recently Capurra *et. al.* [2] have also studied the heart beat data recorded during meditation. They have shown that the embedding of the experimental beat-to-beat interval data acquires a polygonal shape during meditation and that the time course of respiration depends on the meditation technique. They have presented a model for the influence of respiration on the heart beat-to-beat time interval series.

## Meditation data

The term 'meditation' refers to a group of techniques, most of which started in Eastern religious or spiritual traditions. These techniques have been used by many different cultures throughout the world for thousands of years. Today, many people use meditation outside of its traditional religious or cultural settings, for health and wellness purposes. In meditation, a person learns to focus his attention and suspend the stream of thoughts that normally occupy the mind. This practice is believed to result in a state of greater physical relaxation, mental calmness, and psychological balance.

To study the effect of meditation on heart beat we used time series, from physionet [3], of 8 subjects prior and during Chinese Chi meditation. The heart beat-to-beat intervals were obtained from electrocardiogram (ECG) data by Peng et al. [1] and made available in physionet. These eight Chi meditators, 5 women and 3 men (age range 26–35 years), wore a Holter recorder for approximately 10 h during which time they went about their ordinary daily activities. After approximately 5 h of recording, each one of them practiced 1 h of meditation. During these sessions, the Chi meditators sat quietly, listening to the taped guidance of the master. The meditators were instructed to breath spontaneously while visualizing the opening and closing of a perfect lotus on the abdomen.

## Scaling Analysis

Scaling as a manifestation of the underlying dynamics is familiar throughout physics. It has been instrumental in helping scientists gain deeper insights into problems ranging across the entire spectrum of science and technology. Scaling laws typically reflect underlying generic features and physical principles that are independent of detailed dynamics or characteristics of particular models. Scale invariance has been found to hold empirically for a number of complex systems, and the correct evaluation of the scaling exponents is of fundamental importance in assessing if universality classes exist [4]. Scale invariance seems to be widespread in natural systems. Numerous examples of scale invariance properties can be found in the literature like earthquakes, clouds, networks etc [5-7]. To study the effect of meditation on the scaling behavior of heart beat the following analyses were performed.

**Detrended Fluctuation Analysis**

Heartbeat time series is often highly non-stationary. This creates an immediate problem to researchers applying the conventional techniques of time series analysis to heartbeat data analysis. There are various methods of scaling analysis. In recent years, the detrended fluctuation analysis (DFA) proposed by Peng et. al. [8] has been established as an important tool for the detection of long-range correlations in time series with non-stationarities. It avoids the spurious detection of apparent long-range correlations that are an artifact of non-stationarity [9]. This method has been validated on synthetically generated time series superposed with non-stationary trends and also been successfully applied to such diverse field of interest as DNA, heart rate dynamics, temperature fluctuations, rainfall records [8, 10-12] etc.

The DFA procedure consists of the following steps. In the first step, the profile

$$Y(k) = \sum_{i=1}^{k} \left[ x(i) - \langle x \rangle \right] \qquad (1)$$

is determined from the time series $x(i)$ of length N. $\langle x \rangle$ indicates the mean value of $x(i)$'s. Next, the profile $Y(k)$ is divided into $N_n = [N/n]$ non-overlapping segments of equal length $n$. In the next step, the local trend for each segment is calculated by a least-square fit of the data. The y-coordinate of the fitted line is denoted by $Y_n(k)$. Then the detrended time series for the segment duration $n$ is defined as

$$Y_s(k) = Y(k) - Y_n(k) \qquad (2)$$

The root-mean square fluctuation of the original time series and the detrended time series is calculated by

$$F(n) = \sqrt{\frac{1}{N}\sum_{k=1}^{N}\left[Y(k) - Y_n(k)\right]^2} \qquad (3)$$

Repeating this calculation over all segment sizes, a relationship between $F(n)$ and $n$ is obtained. If $F(n)$ behaves as a power law function of $n$, data present scaling:

$$F(n) \propto n^{\beta} \qquad (4)$$

Finally the double logarithmic plot of $F(n)$ versus $n$ is used to calculate the slope, which gives the scaling exponent $\beta$.

If $0 < \beta < 0.5$, the time series is long-range anti-correlated; if $\beta > 0.5$, the time series is long-range correlated. $\beta = 0.5$ corresponds to Gaussian white noise, while $\beta = 1$ indicates the $1/f$ noise, typical of systems in a SOC state. For $\beta > 1$, correlations exist but not like a power law form. $\beta = 1.5$ indicates Brown noise, which is simply the integration of white noise.

**Diffusion Entropy Analysis**

The diffusion entropy analysis (DEA) [13] is primarily developed to detect the scaling behavior of stationary time series. To apply DEA to the heart beat time series, following the idea of Ghasemi et. al. [14], we have calculated the return from the heart beat data using the formula: $\xi_i = \ln(x_{i+1} / x_i)$. The return series is stationary, which can be straightforwardly verified.

DEA is based on the prescription that numbers in a time series $\{\xi_i\}$ are the fluctuations of a diffusion trajectory; see Refs. [13,15,16] for details. Therefore, we shift our attention from the time series $\{\xi_i\}$ to probability density function (pdf) $p(z,t)$ of the corresponding diffusion process. Here $z$ denotes the variable collecting the fluctuations and is referred to as the diffusion variable. The scaling property of $p(z,t)$ takes the form

$$p(z,t) = \frac{1}{t^{\delta}} F\left(\frac{z}{t^{\delta}}\right) \qquad (5)$$

The DEA was developed [13] as an efficient way to detect the scaling and memory in time series for variables in complex systems. This procedure has been successfully applied to sociological, astrophysical, metallurgical and financial time series [17-20]. DEA focuses on the scaling exponent δ evaluated through the Shannon entropy *S(t)* of the diffusion generated by the fluctuations $\{\xi_i\}$ of the time series [13, 15]. Here, the pdf of the diffusion process, $p(z,t)$, is evaluated by means of the sub trajectories $z_n(t) = \sum_{i=0}^{t} \xi_{i+n}$ with $n$=0,1,.. If the scaling condition of Eq. (5) holds true, it is easy to prove that the entropy

$$S(t) = -\int_{-\infty}^{\infty} p(z,t) \ln[p(z,t)] dz \qquad (6)$$

increases in time as

$$S(t) = A + \delta \ln(t) \qquad (7)$$

with

$$A = -\int_{-\infty}^{\infty} dy F(y) \ln[F(y)] = \text{Constant}, \qquad (8)$$

where $y = \frac{z}{t^{\delta}}$. Eq. (7) indicates that in the case of a diffusion process with a scaling pdf, its entropy $S(t)$ increases linearly with $\ln(t)$. The scaling exponent $\delta$ is evaluated from the gradient of the fitted straight line in the linear-log plot of $S(t)$ against $t$.

## Complexity Measure

A system consisting of many parts which are connected in a nonlinear fashion is designated as the complex system. In a complex system, the interaction between the parts allows the emergence of global behavior that would not be anticipated from the behavior of components in isolation, which posses a real threat to deal it properly whereas understanding its behavior offers possibility of spectacular and unforeseen advances in many areas of science and its application. It is this threat and this promise that is making the science of complex systems as of the fastest growing areas of science at the present time. Recently there have been some attempts to quantify the complexity of the complex systems [21-23]. Though there is no formal definition of complexity, it is considered to be a measure of the inherent difficulty to achieve the desired understanding. Simply stated, the complexity of a system is the amount of information necessary to describe it.

Quantifying the complexity of the time series from a physical process may be of considerable interest due to its potential application in evaluating a dynamical model of the system. In this section, we present a quantitative study of the measure of complexity of the heartbeat data recorded during and prior meditation. Entropy based algorithm are often used to quantify the regularity of a time series [24]. Increase in entropy corresponds to the increase in the degree of disorder and for a completely random system it is maximum. Traditional algorithms are single-scale based [21-24]. However, time series derived from the complex systems are likely to present structure on multiple temporal scales. In contrast, time series derived from a simpler system are likely to present structures on just a single scale. For these reasons the traditional single scale based algorithms often yield misleading quantifications of the complexity of a system.

Recently Costa et al. [25-27] introduced a new method, Multiscale Entropy (MSE) analysis for measuring the complexity of finite length time series. This method measures complexity taking into account the multiple time scales. This computational tool can be quite effectively used to quantify the complexity of a natural time series. The first multiple scale measurement of the complexity was proposed by Zhang [28]. Zhang's method was based on the Shannon entropy which requires a large number of almost noise free data. On the contrary, the MSE method uses Sample Entropy (SampEn) to quantify the regularity of finite length time series. SampEn is largely independent of the time series length when the total number of data points is larger than approximately 750 [29]. Thus MSE proved to be quite useful in analyzing the finite length time series over the Zhang's method. Recently MSE has been successfully applied to quantify the complexity of many Physiologic and Biological signals [25-27].

**Multiscale Entropy Analysis**

The MSE method is based on the evaluation of SampEn on the multiple scales. The prescription of the MSE analysis is: given a one-dimensional discrete time series, $\{x_1,.....,x_i,....,x_N\}$, construct the consecutive coarse-grained time series, $\{y^{(\tau)}\}$, determined by the scale factor, $\tau$, according to the equation:

$$y_j^{\tau} = 1/\tau \sum_{i=(j-1)\tau+1}^{j\tau} x_i \qquad (9)$$

where $\tau$ represents the scale factor and $1 \le j \le N/\tau$. The length of each coarse-grained time series is $N/\tau$. For scale one, the coarse-grained time series is simply the original time series. Next we calculate the SampEn for each scale using the following method. Let $\{X_i\} = \{x_1, ....., x_i, ....., x_N\}$ be a time series of length N. $u_m(i) = \{x_i, x_{i+1}, ....., x_{i+m-1}\}, 1 \le i \le N - m$ be vectors of length *m*. Let $n_{im}(r)$ represent the number of vectors $u_m(j)$ within distance *r* of $u_m(i)$, where *j* ranges from 1 to (N-m) and $j \neq i$ to exclude the self matches. $C_i^m(r) = n_{im}(r)/(N - m - 1)$ is the probability that any $u_m(j)$ is within *r* of $u_m(i)$.We then define

$$U^m(r) = 1/(N - m)\sum_{i=1}^{N-m} \ln C_i^m(r) \qquad (10)$$

The parameter Sample Entropy (SampEn) [29] is defined as

$$SampEn(m, r) = \lim_{N \to \infty} \left\{ -\ln \frac{U^{m+1}(r)}{U^m(r)} \right\} \qquad (11)$$

For finite length N the SampEn is estimated by the statistics

$$SampEn(m, r, N) = -\ln \frac{U^{m+1}(r)}{U^m(r)} \qquad (13)$$

Advantage of SampEn is that it is less dependent on time series length and is relatively consistent over broad range of possible r, m and N values. We have calculated SampEn for all the studied data sets with the parameters m=2 and r= 0.15×SD (SD is the standard deviation of the original time series).

**Recurrence Analysis**

Eckman, Kamphorst and Ruelle [30] proposed a new method to study the recurrences and nonstationary behaviour occurring in dynamical system. They designated the method as “recurrence plot” (RP). The method is found to be efficient in identification of system properties that cannot be observed using other conventional linear and nonlinear approaches. Moreover, the method has been found very useful for analysis of nonstationary system with high dimension and noisy dynamics. The method can be outlined as follows: given a time series $\{x_i\}$ of N data points, first the phase space vectors $u_i=\{x_i, x_{i+\tau}, \ldots\ldots, x_{i+(d-1)\tau}\}$are constructed using Taken’s time delay method. The embedding dimension ($d$) can be estimated from the false nearest neighbor method. The time delay ($\tau$) can be estimated either from the autocorrelation function or from the mutual information method. The main step is then to calculate the N×N matrix

$$R_{i,j} = \Theta(\varepsilon_i - \left\| \overrightarrow{x_i} - \overrightarrow{x_j} \right\|, \qquad i,j=1,2,\ldots,N \qquad (14)$$

where $\varepsilon_i$ is a cutoff distance, ||..|| is a norm ( we have taken the Euclidean norm), and $\Theta(x)$ is the Heavyside function. The cutoff distance $\varepsilon_i$ defines a sphere centered at $\overrightarrow{x_i}$. If $\overrightarrow{x_j}$ falls within this sphere, the state will be close to $\overrightarrow{x_i}$ and thus $R_{i,j} = 1$. The binary values in $R_{i,j}$ can be simply visualized by a matrix plot with color black (1) and white (0). This plot is called the recurrence plot.

However, it is often not very straight forward to conclude about the dynamics of the system from the visual inspection of the RPs. Zbilut and Webber [31,32] developed the so called recurrence quantification analysis (RQA) to provide quantification of important aspects revealed through the plot. The RQA proposed by Zbilut and Webber is mostly based on the diagonal structures in the RPs. They defined different measures, the recurrence rate (REC) measures the fraction of black points in the RP, the determinism (DET) measure of the fraction recurrent points forming the diagonal line structure, the maximal length of diagonal structures ($L_{max}$), the entropy (Shannon entropy of the line segment distributions) and the trend (measure of the paling of recurrent points away from the central diagonal). These parameters are used to detect the transitions in the time series. Recently Gao [33] emphasized the importance of the vertical structures in RPs and introduced a recurrence time statistics corresponding to the vertical structures in RP. Marwan et. al. [34] extended Gao's view and defined measures of complexity based on the distribution of the vertical line length. They introduced three new RP based measures: the laminarity, the trapping time (TT) and the maximal length of the vertical structures ($V_{max}$). Laminarity is analogous to DET and gives the measure of the amount of vertical structure in the RP and represents laminar states in the system. TT contains information about the amount as well as the length of the vertical structure. Applying these measures to the logistic map data they found that in contrast to the conventional RQA measures, their measures are able to identify the laminar states i.e. chaos-chaos transitions. The vertical structure based measures were also found very successful to detect the laminar phases before the onset of life-threatening ventricular tachyarrhythmia. Here we have applied the measures proposed by Marwan *et. al.* along with the traditional measures to find the effect of meditation on the heart-rate variability data.

## Results and discussions

Heart beat sequences have been a topic of many time series analysis in the last few years. Recent studies reveal that at normal conditions, heart beat rate fluctuations display extended correlation of the type typically exhibited by dynamical systems far from equilibrium. Scaling analysis of the heart beat sequences showed that certain disease states alter the scale-invariant correlation property of the normal heart beat [35]. Here we study how meditation affects the scaling behavior of the heart beat sequences. Fig. 3 shows the typical plot DFA of the representative data of Fig. 1. As mentioned earlier, DFA reveals the correlation property of the time series data. It is seen that meditation severely affects the long range correlation property of the heart beat. There is cross over in the scaling behavior. The scaling exponents are shown in the figure. DEA is relatively a new method to detect the scaling behavior of time series data. In contrast to several variance-based methods, including DFA, DEA studies the scaling behavior of the probability distribution. This makes it useful to detect the scaling behavior of time series even if the variance diverges. Due to the stationary requirement, we have applied DEA to the return series. Fig 4 shows the typical plot of the DEA of the return of the Fig 2. The inset in Fig 4 shows the plot DEA of the heart beat data of Fig 1. The DEA of return shows that during meditation the long range correlation in heart beat sequences is lost. Moreover, DEA reveals periodic behavior in the heart beat return during mediation.

Quantification of complexity of physiologic time series, primarily of heart rate, has been of considerable interest during recent years. Costa et. al. [25] applied MSE to quantify the complexity of heart rate time series derived from healthy subjects, patients with severe congestive heart failure and patients with the cardiac arrhythmia, atrial fibrillation. Here we apply MSE to the heart beat rate time series to study the complexity of the heart rate during meditation. Fig 5 shows the variation of the sample entropy as a function of scale factor for the heart rate time series recorded prior and during meditation. It is seen that at small scale the sample entropy is almost similar for both cases but at larger scale the sample entropy is much higher for the heart beat rate time series recorded during meditation. This signifies that the complexity of the heart beat rate is higher during meditation.

RQA has been used to study the heart rate variability data on many occasions [36-38]. The false nearest neighbor method yielded embedding dimension ~ 10. The time delay obtained from the mutual information is ~ 10. We have used $d$ and $\tau$ both 10 in the RQA. We have selected the value 2 for $\varepsilon_i$, from the scaling curve of REC vs. $\varepsilon_i$ as suggested in literature [36]. Figure 6 shows the typical recurrence plots for the pre and during mediation heart rate time series data. We have calculated the RQA parameters for several small time window (epoch) of the heart beat time series. Figure 7 shows the variation of different RQA parameters with the epoch number. The differences in the RQA parameters for pre and during meditation hear beat data is evident from the plots. To confirm the difference in the RQA parameters we have carried out nonparametric Mann-Whitney U test [39] on the parameters. The results of the test are shown in Table 1.

The results in this paper raise several questions on the effect of meditation on heart as well as the origin of scaling behavior in the Heart rate variability. The answers to the questions are not straightforward-further study is necessary which will in turn also help to understand the physiology of heart.

## Conclusion

In conclusion, we have studied the effect of meditation on human heart beat by analyzing the heart beat record from eight samples prior and during Chi meditation. DFA revealed that mediation destroys the long range correlation exhibited by the normal heart. DEA of the heart beat return showed a periodic behavior in the heart beat during meditation. From the MSE analysis it is found that the complexity of the heart rate variability increases during meditation. The recurrence analysis revealed the differences in the heart rate variability data recorded during and before meditation.

Table 1. Various RQA parameters before and during meditation and nonparametric Mann-Whitney U-test: p represents significance, * is for p<0.001 and ** is for p<0.0001.

| RQA parameter | Before meditation | After meditation | p-value |
|---|---|---|---|
| REC | 0.037±0.021 | 0.008±0.001 | ** |
| DET | 0.503±0.104 | 0.679±0.027 | ** |
| $L_{max}$ | 129.2±100.4 | 298.7±96.4 | ** |
| $V_{max}$ | 27.1±36.4 | 7.4±1.3 | ** |
| Laminarity | 0.69±0.06 | 0.79±0.01 | ** |
| TT | 2.98±0.55 | 2.57±0.08 | * |

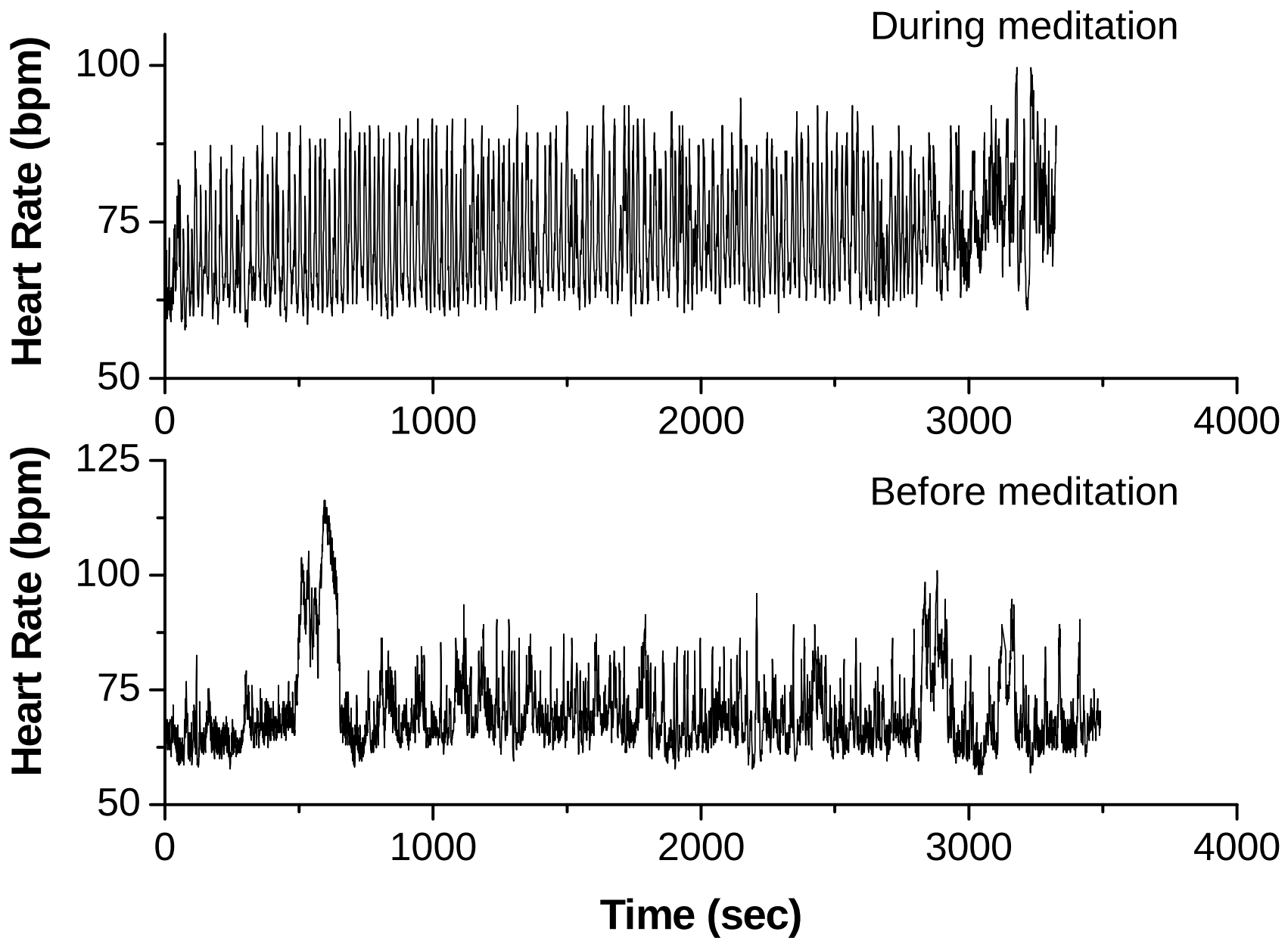

Figure 1 Representative instantaneous heart rate time series before and during meditation

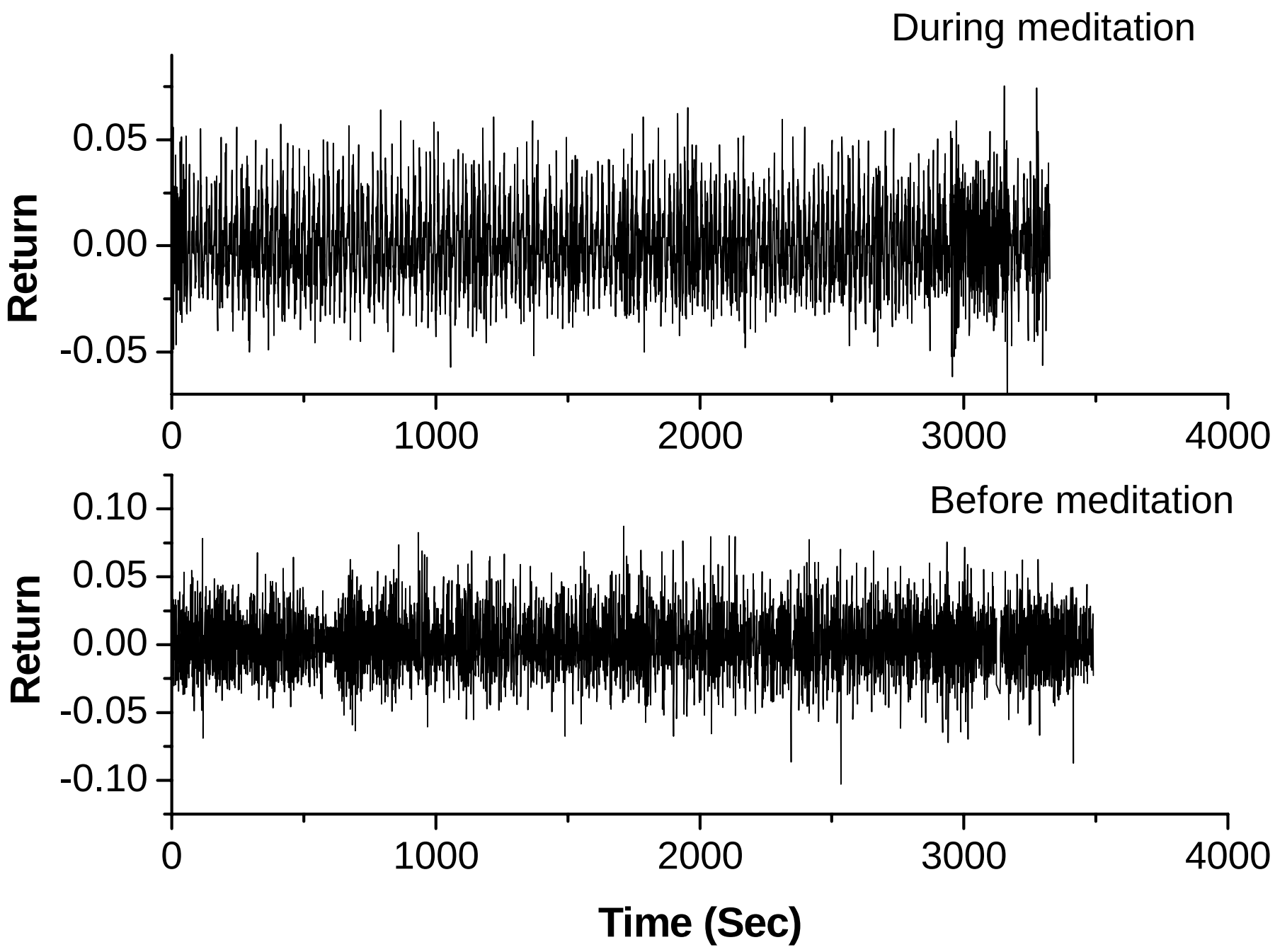

Figure 2 Return of the representative instantaneous heart rate time series before and during meditation

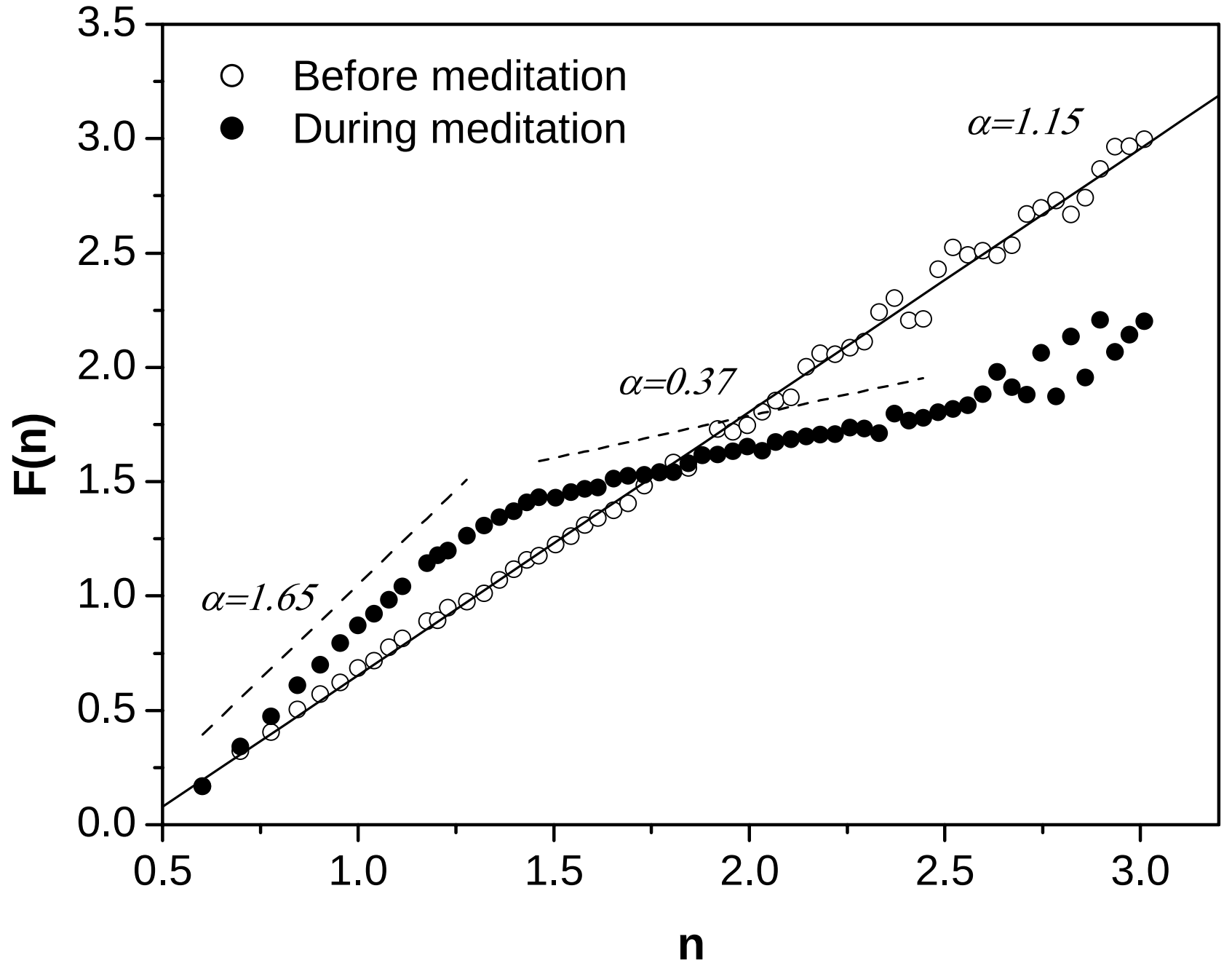

Figure 3 Detrended Fluctuation Analysis of the representative instantaneous heart rate time series before and during meditation

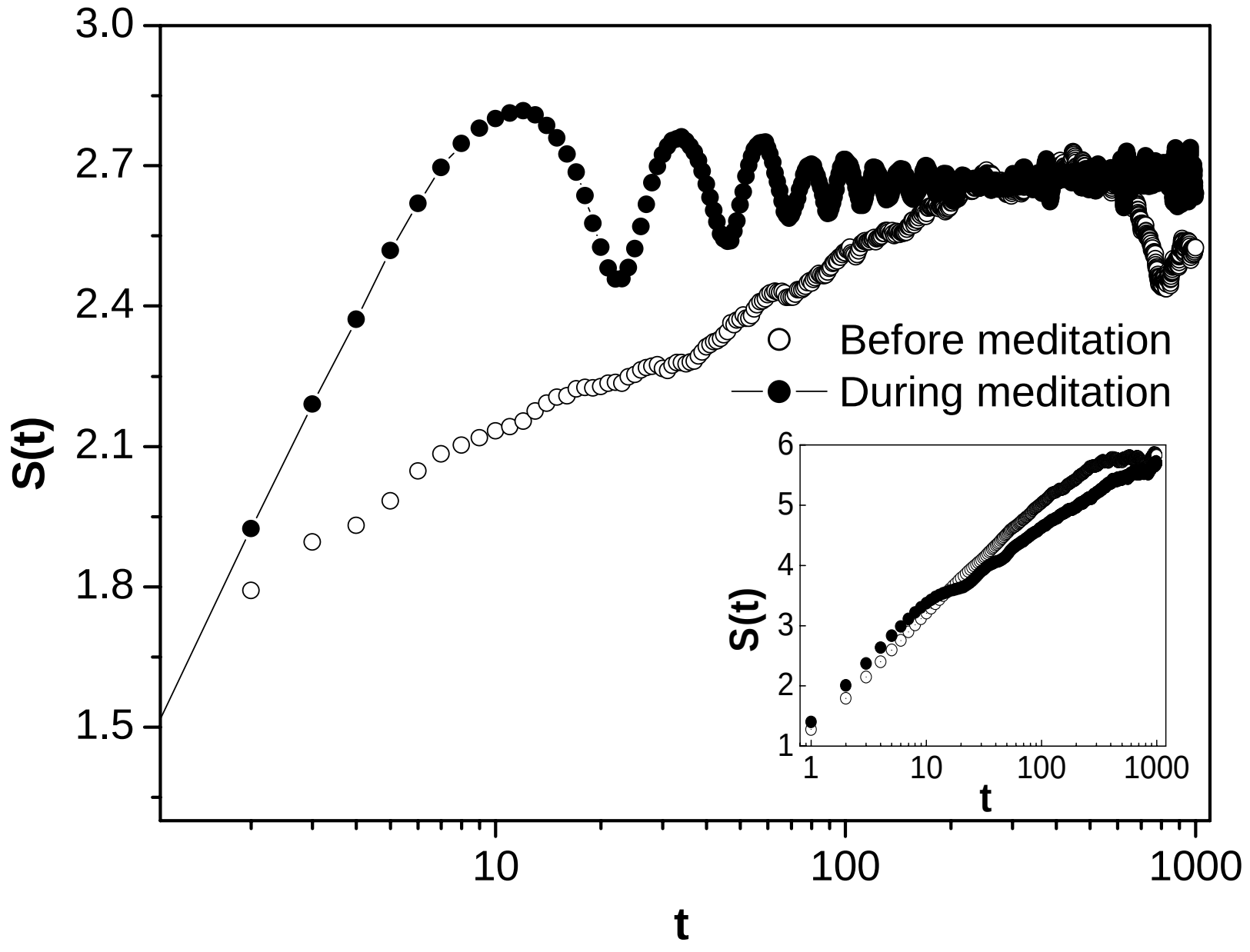

Figure 4 Diffusion Entropy Analysis of the return of the representative instantaneous heart rate time series before and during meditation. The inset shows the DEA of the heart rate time series.

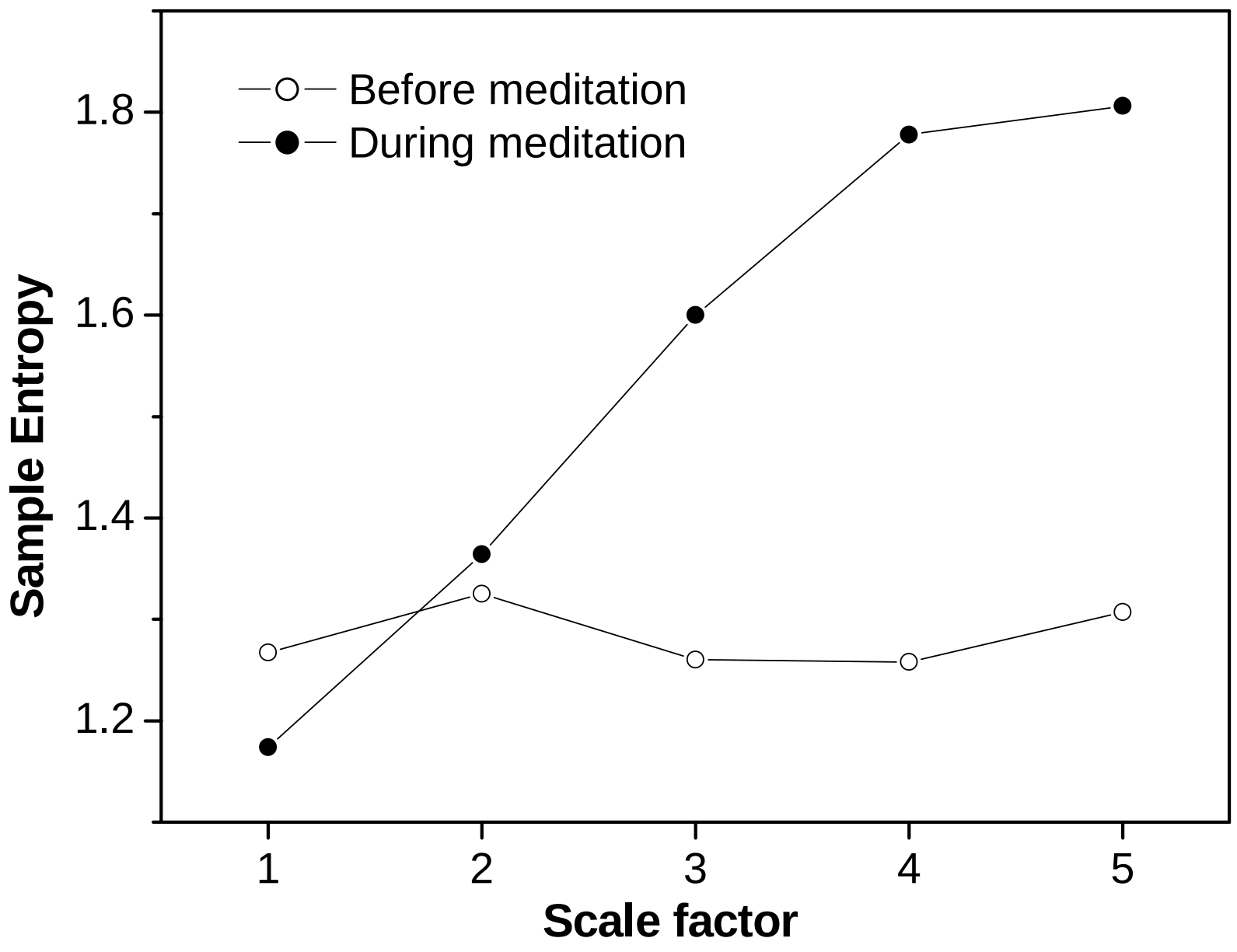

Figure 5 Multiscale Entropy Analysis of the representative instantaneous heart rate time series before and during meditation

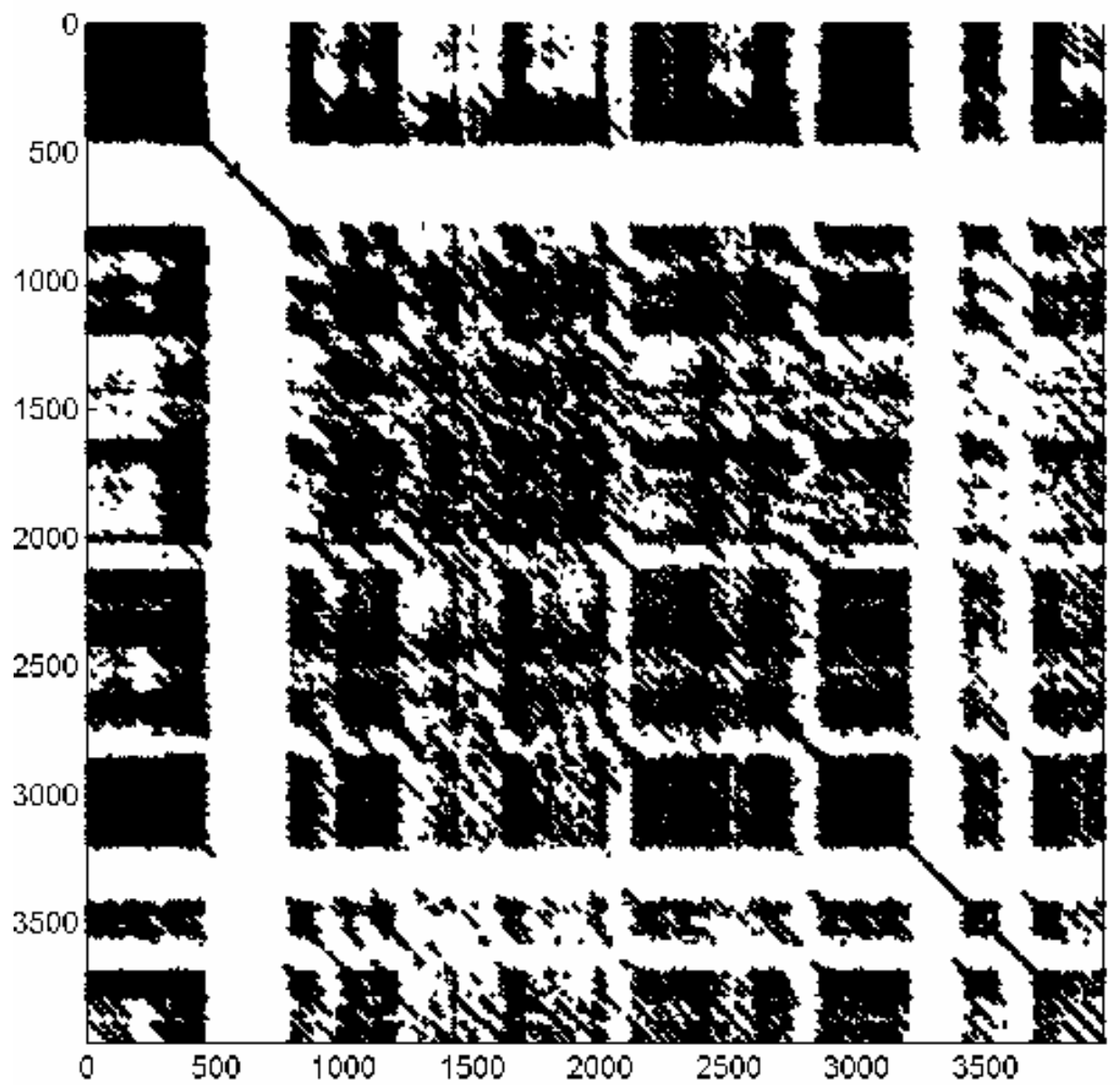

(a)

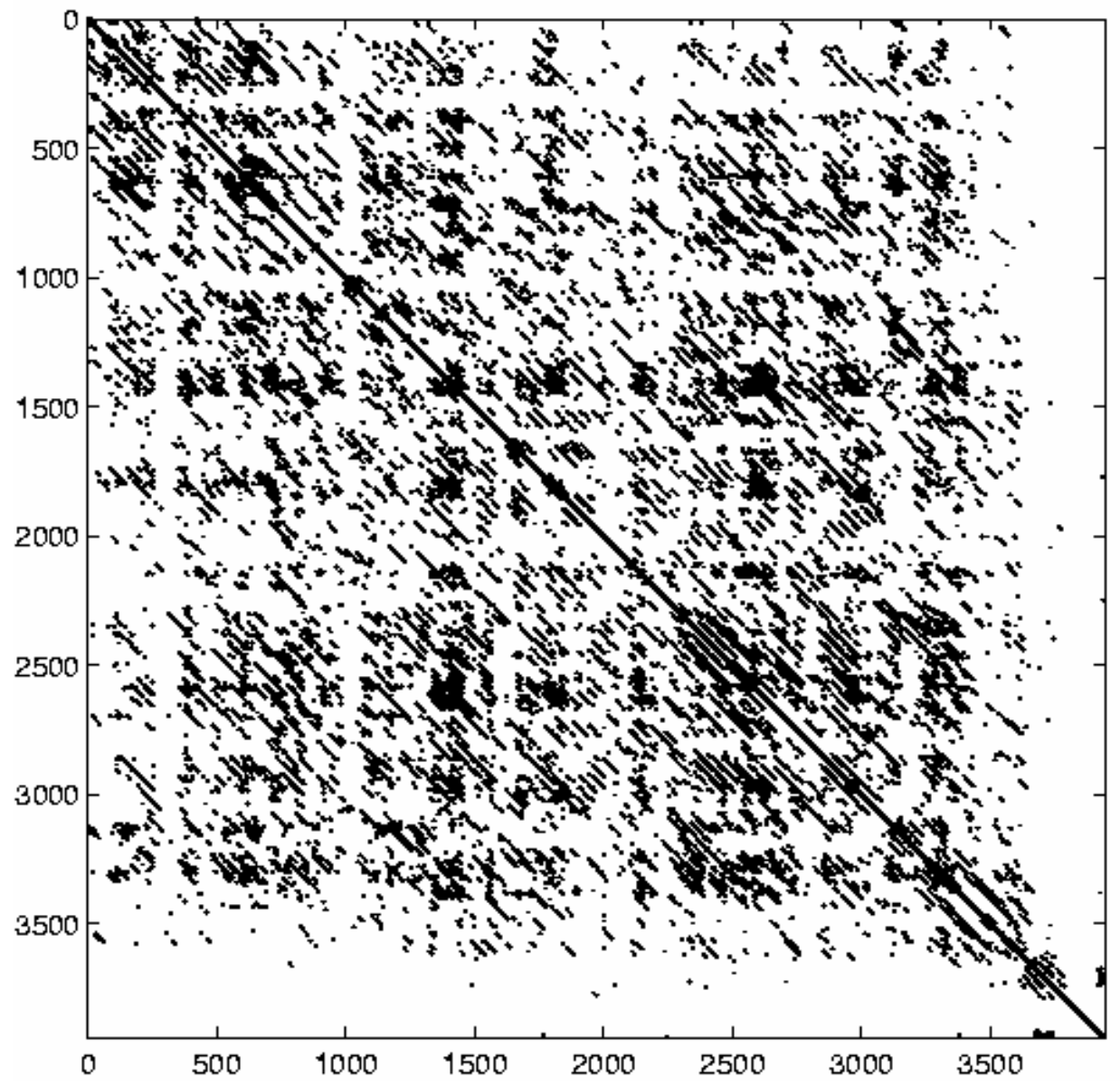

(b)

Figure 6 Recurrence plots of the representative instantaneous heart rate time series (a) before and (b) during meditation

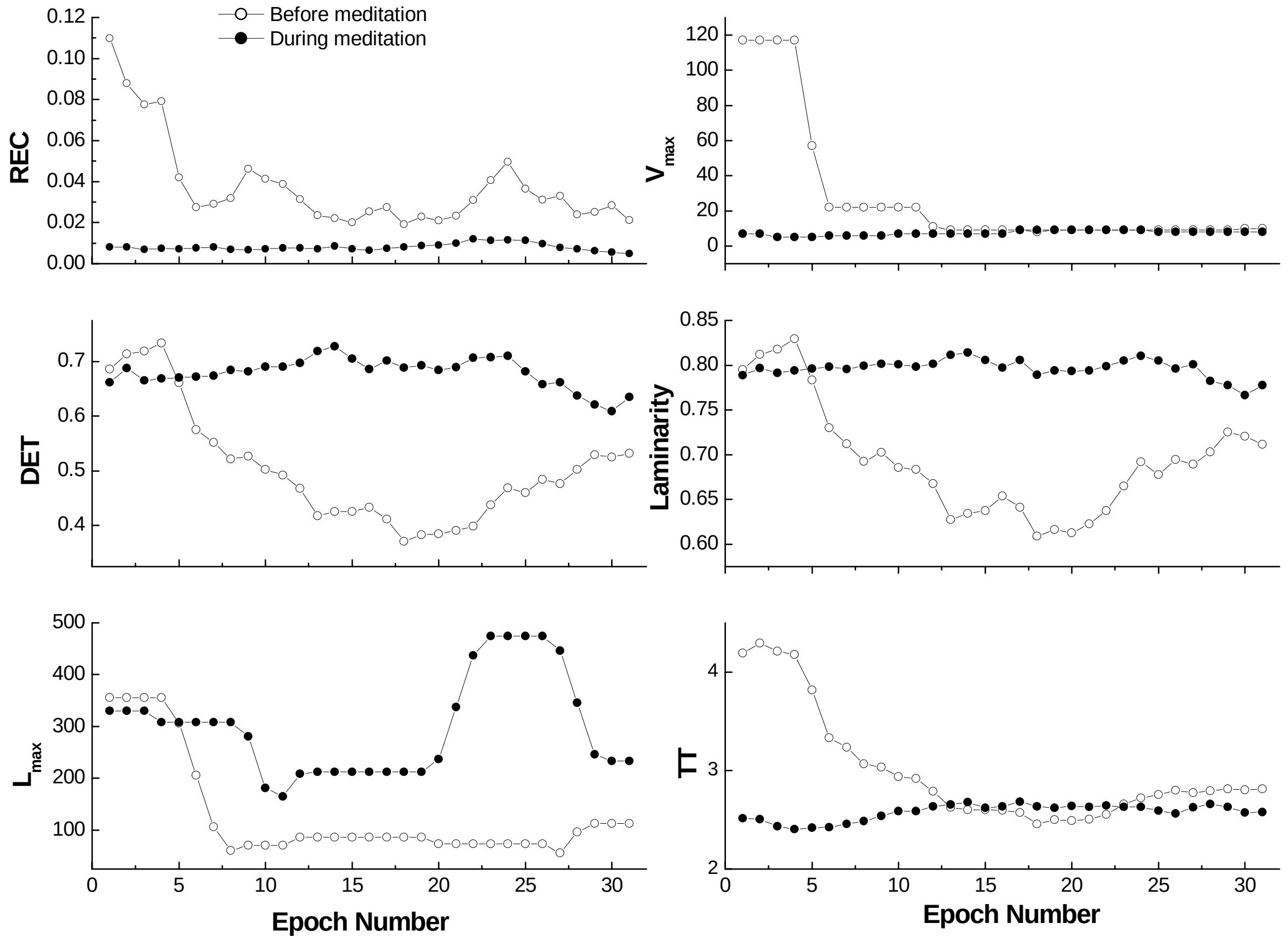

Figure 7 RQA parameters of the representative instantaneous heart rate time series before and during meditation